\documentclass[copyright,creativecommons]{eptcs}
\providecommand{\event}{HPC'10}  
\providecommand{\firstpage}{1}
\usepackage{breakurl}
\usepackage{bm,graphicx,amsbsy,amsmath,amsfonts,amsthm,color,mathrsfs}

\begin{document}

\title{Fast and robust two- and three-qubit swapping gates on multi-atomic ensembles in quantum electrodynamic cavity}

\author{Sergey N. Andrianov
\institute{Institute for Informatics of Tatarstan Academy of
Sciences, 20 Mushtary, Kazan, 420012, Russia} \institute{Physical
Department of Kazan State University, Kremlevskaya 18, Kazan,
420008, Russia}
\email{andrianovsn@mail.ru}
\and
Sergey A. Moiseev
\institute{Institute for Informatics of
Tatarstan Academy of Sciences, 20 Mushtary, Kazan, 420012, Russia}
\institute{Kazan Physical-Technical
Institute of the Russian Academy of Sciences, 10\/7 Sibirsky Trakt,
Kazan, 420029, Russia}
\institute{Physical Department of Kazan State University,
Kremlevskaya 18, Kazan, 420008, Russia}
\email{samoi@yandex.ru}
}

\def\titlerunning{Solid state multi-ensemble quantum computer}
\def\authorrunning{S.N. Andrianov \& S.A. Moiseev}

\maketitle

\begin{abstract}
Qubits on multi-atomic ensembles in a common optical resonator are considered. With that, possible constructions of swap, square root of swap and controlled swap quantum gates are analyzed. Dynamical elimination of excess quantum state and collective blockade mechanism are proposed for realizations of the two and three qubit gates.
\end{abstract}

\section{Introduction}
\label{sec:introduction}
The creation of a quantum computer is an outstanding fundamental and practical problem. The quantum computer could be used for the execution of very complicated tasks which are not solvable with the classical computers. The first prototype of a solid state quantum computer was created in 2009 with superconducting qubits \cite{DiCarlo2009}. However, it suffers from the decoherent processes and it is desirable to find more practical encoding of qubits with long-lived coherence. It could be single impurity or vacancy centers in solids \cite{Balasubramanian2009} but their interaction with electromagnetic radiation is rather weak.  So, here, ensembles of atoms were proposed for the qubit encoding by using the dipole blockade mechanism in order to turn multilevel systems in two level ones \cite{Saffman2010}. But dipole-dipole based blockade introduces an additional decoherence that limits its practical significance. Recently, the collective blockade mechanism has been proposed for the system of three-level atoms by using the different frequency shifts for the Raman transitions between the collective atomic states characterized by a different number of the excited atoms \cite{Shahriar2007}. Here, we propose a two qubit gate by using another collective blockade mechanism in the system of two level atoms based on exchange interaction via the virtual photons between the multi-atomic ensembles in the resonator. Also we demonstrate the possibility of a three qubit gate (Controlled SWAP gate) using a suppression of the swap-process between two multi-atomic ensembles due to dynamical shift of the atomic levels controlled by the states of photon encoded qubit.

\section{Swap gates}
\label{sec:swap}
\begin{figure}
  \includegraphics[width=0.6\textwidth,height=0.4\textwidth]{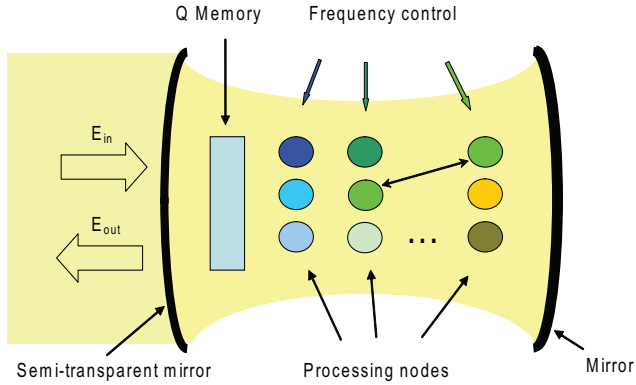}
  \caption{Layout for realization of two-qubit $iSWAP$ gates. Input and output quantum information is encoded through the quantum states of the in-coming ($E_{in}$) and out-coming ($E_{out}$) single photon fields.}
  \label{Figure1}
\end{figure}

Let us consider a plurality of the atomic systems (nodes) situating in a common electrodynamic resonator with a quantum memory (QM) node as depicted in Fig.
\ref{Figure1}.
For realization of two-qubit gates we transfer the two qubits from QM node to the 1-st and 2-rd processing nodes and equalize the carrier frequencies of the nodes at time moment t=0 with some detuning from the resonator mode frequency
$\omega_1-\omega_0=\Delta_{1}=\omega_2-\omega_0=\Delta_{2}=\Delta$.
It yields to the following initial state of 1-st and 2-nd nodes in the interaction picture

\begin{align}
\label{eq1}
\psi_{in}(0)=\{\alpha_1|0>_1+\beta_1|1>_1\}\{\alpha_2|0>_2+\beta_2|1>_2\},
\end{align}

\noindent
where
$|\alpha_{1,2}|^2+|\beta_{1,2}|^2=1$.
Here, we have introduced the following states:
$|0>_m=|0_1,0_2,...,0_{N_m}>$
corresponding to the ground state of the m-th node,
$|1>_m=\sqrt{1/N}\sum_{j}^{N_m}|0>_1|0>_2...|1>_j...|0>_{N_m}$
and
$|2>_m=\sqrt{2/N(N-1)}\sum_{i\neq j}^{N_m}|0>_1|0>_2...|1>_i...|1>_j...|0>_{N_m}$
are the collective states of m-th node with single and two atomic excitations.
Equal frequencies of the two nodes results in interaction of the atoms via the virtual processes of resonant circuit quanta
determined by effective Hamiltonian1,2 between these nodes.
In order to obtain this interaction, we start from initial Hamiltonian
$\hat H=\hat H_0+\hat H_1$
where
$\hat H_0=\hat H_a+\hat H_r$
is main Hamiltonian and
$\hat H_1=\hat H_{r-a}$
is perturbation Hamiltonian. Here,
$\hat H_a=\hat H_{a_1}+\hat H_{a_2}$
is Hamiltonian of atoms in nodes 1 and 2 and
$\hat H_r$
is Hamiltonian of photons. With that,
$\hat H_{a_1}=\hbar \omega_0 \sum_{j_1}S_{j_1}^z$
and
$\hat H_{a_2}=\hbar \omega_0 \sum_{j_2}S_{j_2}^z$
where
$\omega_0$
is the frequency of working transitions in atoms,
$S_{j_1}^z$
and
$S_{j_2}^z$
operators of effective spin ½  z-projection in two-level model for atoms in sites
$j_1$
and
$j_2$
of nodes 1 and 2;
$\hat H_r=\hbar \omega_{k_0}a_{k_0}^+a_{k_0}$
where
$\omega_{k_0}$
is frequency of photons with wave vector
$k_0$,
$a_{k_0}^+$
and
$a_{k_0}$
are creation and annihilation operators for photons. We have for interaction of atoms with photons
$\hat H_{r-a}=H_{r-a}^{(1)}+H_{r-a}^{(1)}$
in nodes 1 and 2 the following expressions:
\begin{align}
\label{eq2}
H_{r-a}^{(1)}=\sum_{j_1}(g_{k_0}^{(1)}e^{ik_0r_{j_1}}S_{j_1}^+a_{k_0}+g_{k_0}^{(1)*}e^{-ik_0r_{j_1}}S_{j_1}^-a_{k_0}^+),
\end{align}

\begin{align}
\label{eq3}
H_{r-a}^{(2)}=\sum_{j_2}(g_{k_0}^{(2)}e^{ik_0r_{j_2}}S_{j_2}^+a_{k_0}+g_{k_0}^{(2)*}e^{-ik_0r_{j_2}}S_{j_2}^-a_{k_0}^+),
\end{align}

\noindent
where $g_{k_0 }^{\left( \alpha \right)} $ are interaction constants, $S_{j_2
}^ + $ are raising and lowering operators for spin
$\raise.5ex\hbox{$\scriptstyle 1$}\kern-.1em/
\kern-.15em\lower.25ex\hbox{$\scriptstyle 2$} $ in two level model, $\vec
{r}_{j_\alpha } $ are radius vectors for atoms in sites $j_\alpha $ of nodes
$\alpha = 1,2$.

We perform unitary transformation of Hamiltonian $H_s = e^{ - s}He^s$ that
yields in the second degree on small perturbation the following result:

\begin{equation}
\label{eq2-4}
H_s = H_0 + \frac{1}{2}\left[ {H_1 ,s} \right],
\end{equation}

\noindent
when relation $ H_1 + \left[ {H_0 ,s} \right] = 0$ is valid. Using relation (\ref{eq3}) we find $ s = s_1 + s_2 $,

\begin{equation}
\label{eq5}
s_1 = \sum\limits_{j_1 } {\left( {\alpha _1 g_{k_0 }^{\left( 1 \right)}
e^{i\vec {k}_0 \vec {r}_{j_1 } }S_{j_1 }^ + a_{k_0 } + \beta _1 g_{k_0
}^{\left( 1 \right)\ast } e^{ - i\vec {k}_0 \vec {r}_{j_1 } }S_{j_1 }^ -
a_{k_0 }^ + } \right)} ,
\end{equation}

\begin{equation}
\label{eq6}
s_2 = \sum\limits_{j_2 } {\left( {\alpha _2 g_{k_0 }^{\left( 2 \right)}
e^{i\vec {k}_0 \vec {r}_{j_2 } }S_{j_2 }^ + a_{k_0 } + \beta _2 g_{k_0
}^{\left( 2 \right)\ast } e^{ - i\vec {k}_0 \vec {r}_{j_2 } }S_{j_2 }^ -
a_{k_0 }^ + } \right)},
\end{equation}

\noindent
where

\begin{equation}
\label{eq7}
\alpha _{1,2}= -\beta _{1,2}= - \frac{1}{\hbar \left( {\omega _0 - \omega _{k_0 } }
\right)} = - \frac{1}{\hbar \Delta }.
\end{equation}

\noindent
Substituting (\ref{eq5}) and (\ref{eq6}) into (\ref{eq2-4}), we get

\begin{align}
\label{eq9}
 H_s & = \hbar \omega _{k_0 } a_{k_0 }^ + a_{k_0 } + \sum\limits_m^{1,2}
{\sum\limits_{j_m } {\hbar \omega _m S_{j_m }^z } } + 2\sum\limits_m^{1,2}
{\sum\limits_{j_m } {\frac{\left| {g_{k_0 }^{\left( m \right)} }
\right|^2}{\hbar \Delta }a_{k_0 }^ + a_{k_0 } S_{j_m }^z } } +
\sum\limits_m^{1,2} {\sum\limits_{i_m j_m } {\frac{\left| {g_{k_0 }^{\left(
m \right)} } \right|^2}{\hbar \Delta }S_{i_m }^ + S_{j_m }^ - + } }
\nonumber   \\
& + \frac{1}{\hbar \Delta }\sum\limits_{j_1 j_2 } {\left( {g_{k_0 }^{\left( 1
\right)} g_{k_0 }^{\left( 2 \right)\ast } e^{i\vec {k}_0 \vec {r}_{j_1 j_2 }
}S_{j_1 }^ + S_{j_2 }^ - + g_{k_0 }^{\left( 1 \right)\ast } g_{k_0 }^{\left(
2 \right)} e^{ - i\vec {k}_0 \vec {r}_{j_1 j_2 } }S_{j_1 }^ - S_{j_2 }^ + }
\right)}.
\end{align}

The first term is unchanged energy of photons, the second term is unchanged
energy of atoms in nodes 1 and 2, the third term is atomic energy shifts due
to photons, the forth term is atomic intra-node swap energy, the fifth term
is atomic inter-node swap energy.

According to (\ref{eq9}) effective interaction of atoms is $\hat {H}_{eff} =
\sum\nolimits_{m = 1}^2 {\hat {H}_{node}^{(m)} } + \hat {H}_{int} $, where \linebreak
$\hat {H}_{node}^{(m)} = \hbar \Omega _\sigma \sum\nolimits_{i_m j_m }^N
{e^{i\vec {k}_0 \vec {r}_{i_m j_m } }S_{i_m }^ + S_{j_m }^ - } $ is a
long-range spin-spin interaction in m-th node, \linebreak $\hat {H}_{int} = \hbar
\Omega _\sigma \sum\nolimits_{j_1 ,j_2 = 1}^N {\left( {e^{i\vec {k}_0 \vec
{r}_{j_1 j_2 } }S_{j_1 }^ + S_{j_2 }^ - + e^{ - i\vec {k}_0 \vec {r}_{j_1
j_2 } }S_{j_1 }^ - S_{j_2 }^ + } \right)} $ (where $\Omega _\sigma = \vert
g_\sigma \vert ^2 / \Delta )$ describes a spin-spin interaction between the
two nodes ($N_1 = N_2 = N)$,$\vec {k}_0 $ is wave vector of resonant mode.
Let's introduce the collective basis states of the two nodes:$\left| \psi
\right\rangle _1 = \left| 0 \right\rangle _1 \left| 0 \right\rangle _2 $,
$\left| \psi \right\rangle _2 = \left| 1 \right\rangle _1 \left| 0
\right\rangle _2 $, $\left| \psi \right\rangle _3 = \left| 0 \right\rangle
_1 \left| 1 \right\rangle _2 $, $\left| \psi \right\rangle _4 = \left| 1
\right\rangle _1 \left| 1 \right\rangle _2 $ and $\left| \psi \right\rangle
_5 = 1 / \sqrt 2 \{\left| 2 \right\rangle _1 \left| 0 \right\rangle _2 +
\left| 0 \right\rangle _1 \left| 2 \right\rangle _2 \}$. It is important
that the Hamiltonian $\hat {H}_{eff} $ has a matrix representation in the
basis of the five states which is separated from other states of the
multi-atomic system

\begin{equation}
\label{eq10}
\left( {{\begin{array}{*{20}c}
 0 \hfill & 0 \hfill & 0 \hfill & 0 \hfill & 0 \hfill \\
 0 \hfill & {N\Omega _\sigma } \hfill & {N\Omega _\sigma } \hfill & 0 \hfill
& 0 \hfill \\
 0 \hfill & {N\Omega _\sigma } \hfill & {N\Omega _\sigma } \hfill & 0 \hfill
& 0 \hfill \\
 0 \hfill & 0 \hfill & 0 \hfill & {2N\Omega _\sigma } \hfill & {2\Omega
_\sigma \sqrt {N(N - 1)} } \hfill \\
 0 \hfill & 0 \hfill & 0 \hfill & {2\Omega _\sigma \sqrt {N(N - 1)} } \hfill
& {2\Omega _\sigma (N - 1)} \hfill \\
\end{array} }} \right).
\end{equation}

By using (\ref{eq10}), we find the unitary evolution of the atomic systems which
couples independently two pairs of the quantum states $\left| \psi
\right\rangle _2 \leftrightarrow \left| \psi \right\rangle _3 $ and $\left|
\psi \right\rangle _4 \leftrightarrow \left| \psi \right\rangle _5 $

\begin{align}
\label{eq11}
 \Psi _1 \left( t \right) &= \alpha _2 \alpha _3 \psi _1 \mbox{ } + \exp ( -
i\Omega _\sigma Nt)\{\beta _2 \alpha _3 [\cos (\Omega _\sigma Nt)\psi _2 -
i\sin (\Omega _\sigma Nt)\psi _3 ]
\nonumber  \\
&+ \alpha _2 \beta _3 [\cos (\Omega _\sigma Nt)\psi _3 - i\sin
(\Omega _\sigma Nt)\psi _2 ]\}
\nonumber   \\
& + \exp ( - i2\Omega _\sigma Nt)\beta _2 \beta _3 [\cos (2\Omega
_\sigma Nt)\psi _4 - i\sin (2\Omega _\sigma Nt)\psi _5 ],
\end{align}

\noindent
where we have assumed a large number of atoms $N \gg 1$. The solution
demonstrates two coherent oscillations with the frequency $\Omega _\sigma N$
for the first pair $\left| \psi \right\rangle _2 \leftrightarrow \left| \psi
\right\rangle _3 $ and with the double frequency $2\Omega _\sigma N$ for the
second pair $\left| \psi \right\rangle _4 \leftrightarrow \left| \psi
\right\rangle _5 $. The oscillations are drastically accelerated N-times
comparing to the case of two coupled two-level atoms so we can use even bad
common resonator with relatively lower quality factor.

It is known [4,5] that the evolution of the two coupled two level atoms can
lead to $iSWAP$ and $\sqrt {iSWAP} $ gates. The $iSWAP$ and $\sqrt {iSWAP} $
gates work in the Hilbert space of four states $\left| \psi \right\rangle _1
,...,\left| \psi \right\rangle _4 $ and these gates are important for
realization of the complete set of the universal quantum gates \cite{Imamoglu,Schuch}.
$iSWAP$ gate provides exchange of the two quantum states between the two
nodes. In our case we get that $iSWAP$ gate occurs at shortened time
$t_{iSWAP} = \pi / 2\Omega _\sigma N$ sec

\begin{equation}
\label{eq12}
\Psi _1 \left( {t_{iSWAP} } \right) = \{\alpha _2 \left| 0 \right\rangle _1
- \beta _2 \left| 1 \right\rangle _1 \}\{\alpha _1 \left| 0 \right\rangle _2
- \beta _1 \left| 1 \right\rangle _2 \}.
\end{equation}

We also note that by choosing different carrier frequencies we can realize
the described $iSWAP$  operation for many pairs of nodes simultaneously due to
exploitation of the independent virtual quanta for each pair in
the QED cavity. It is interesting that the $iSWAP$  gate provides a perfect
elimination of transfer of the initial state to the state $\left| \psi
\right\rangle _5 $ that occurs only at t=t$_{iSWAP}$.

\section{Square root swap gates}
\label{sec:square root}

The situation is more complicated for realization of $\sqrt {iSWAP} $ gate
because it is impossible to eliminate state $\left| \psi \right\rangle _5 $
with evolution based on matrix (12). Below, we propose a universal mechanism
for \textit{collective dynamical elimination} (CDE --procedure) of the state $\left| \psi \right\rangle _5 $ for
realization of $\sqrt {iSWAP} $ gate by using the multi-atomic ensemble
encoding for single qubits and cavity mediated collective interaction.

\begin{figure}
  \includegraphics[width=0.35 \textwidth,height=0.35 \textwidth]{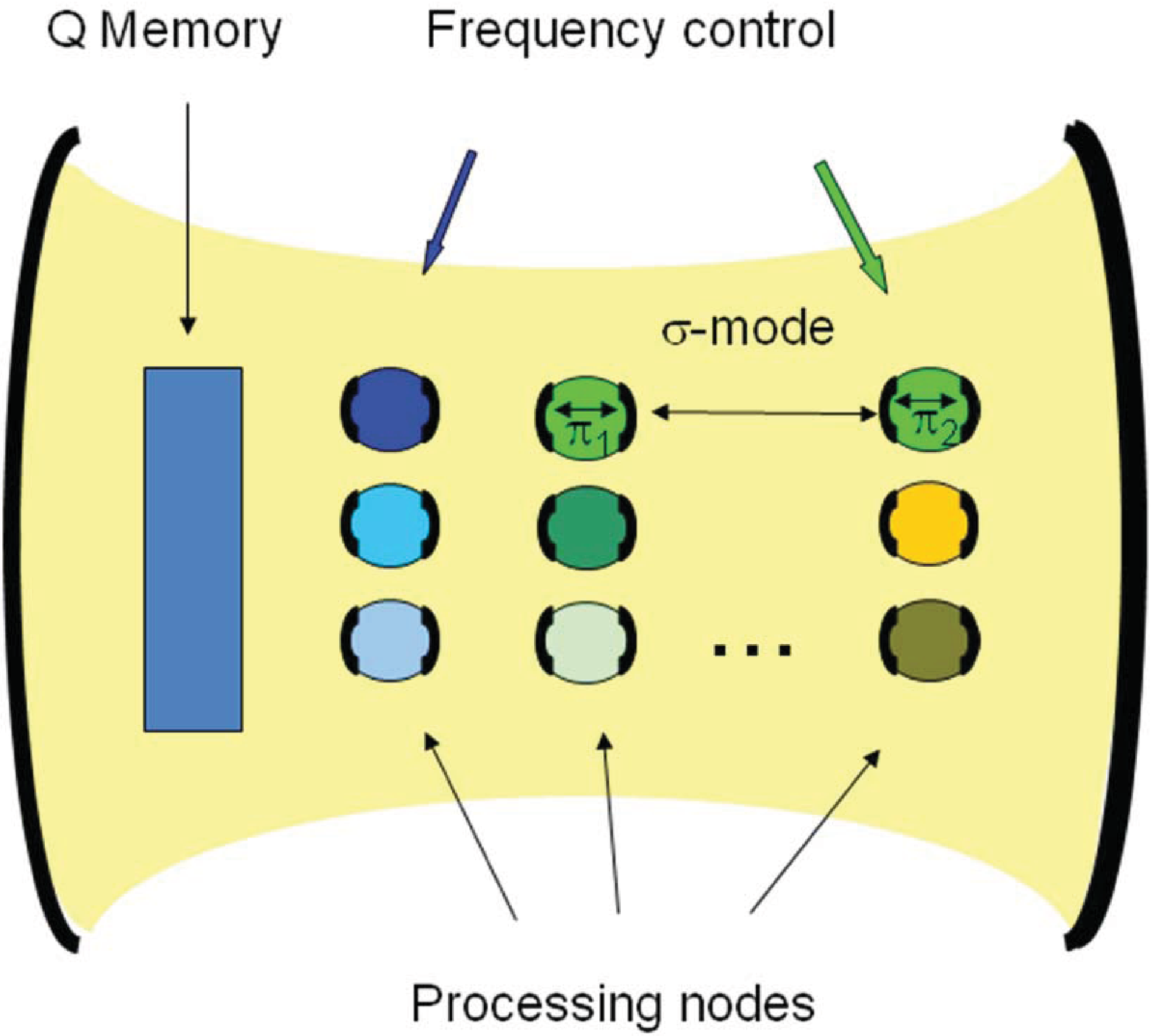}
  \caption{Layout for realization of the two-qubit  $\sqrt{iSWAP}$  gates ($\sigma$ is the mode of common cavity; $\pi_1$  and $\pi_2$  are the local modes).}
  \label{Figure2}
\end{figure}

Scheme of spatial arrangement of the processing nodes and cavities for
realization of the $\sqrt {iSWAP} $ is presented in Fig. \ref{Figure2}.
Here, we insert
the 1-nd and 2-rd nodes in two different single mode QED cavities
characterized by high quality factors for $\pi $-modes. We assume that each
$\pi $-mode interacts only with the atoms of one node and is decoupled from
the basic cavity field mode that is possible for large enough spectral
detuning of the local QED cavity modes.

Thus, we take the additional field Hamiltonians $H_{r_\pi } =
\sum\limits_m^{1,2} {\hbar \omega _{k_0 } a_{k_0 \pi _m }^ + a_{k_0 \pi _m }
} $ and interaction of photons with the atoms in the 1$^{st}$ and the 2$^{nd}$ nodes $\quad H_{r -
a}^{\left( \pi \right)} = \sum\limits_m^{1,2} {\sum\limits_{j_m } {\left(
{g_{k_0 \pi _m }^{\left( m \right)} e^{i\vec {k}_0 \vec {r}_j }S_{j_m }^ +
a_{k_0 \pi _m } + g_{k_0 \pi _m }^{\left( m \right)\ast } e^{ - i\vec {k}_0
\vec {r}_j }S_{j_m }^ - a_{k_0 \pi _m }^ + } \right)} } $. By assuming a
large enough spectral detuning of atomic frequencies from the field mode and
absence of real photons in the QED cavities we find the following effective
Hamiltonian similar to previous section

\begin{align}
\label{eq13a}
 H_s & = \sum\limits_m^{1,2} {\sum\limits_{j_m } {\hbar \omega _m S_{j_m }^z }
} + \sum\limits_m^{1,2} {\sum\limits_{i_m j_m } {\frac{\left| {g_{k_0 \sigma
}^{\left( m \right)} } \right|^2}{\hbar \Delta ^{\prime} _m } e^{i\vec {k}_0 \vec
{r}_{i_m j_m } }S_{i_m }^ + S_{j_m }^ - + \sum\limits_m^{1,2}
{\sum\limits_{i_m j_m } {\frac{\left| {g_{k_0 \pi }^{\left( 1 \right)} }
\right|^2}{\hbar \Delta^{\prime}_m } e^{i\vec {k}_0 \vec {r}_{i_m j_m } }S_{i_m }^
+ S_{j_m }^ - } } + } }
\nonumber \\
& + \frac{1}{2\hbar }\left( {\frac{1}{\Delta _1 } + \frac{1}{\Delta _2 }}
\right)\sum\limits_{j_1 j_2 } {\left( {g_{k_0 \sigma }^{\left( 1 \right)}
g_{k_0 \sigma }^{\left( 2 \right)\ast } e^{i\vec {k}_0 \vec {r}_{j_1 j_2 }
}S_{j_1 }^ + S_{j_2 }^ - + g_{k_0 \sigma }^{\left( 1 \right)\ast } g_{k_0
\sigma }^{\left( 2 \right)} e^{ - i\vec {k}_0 \vec {r}_{j_1 j_2 } }S_{j_2 }^
- S_{j_2 }^ + } \right)},
\end{align}

\noindent
where $\Delta _{1,2} = \omega _{1,2} - \omega _o $ are the atomic frequency
detunings from the common cavity mode and $\Delta^{\prime} _{1,2} = \omega _{1,2} -
\omega _{k_o } $ are the atomic detunings from the frequency of the local
QED cavities having the same frequency $\omega _{k_o } $. To be concrete, we
take below $\Delta _{1,2}^{\prime} = - \Delta _{1,2} = - \Delta $, $\Delta > 0$.

The second and third terms in Eq. (\ref{eq13a}) describes the atom-atom interactions
inside each node via the exchange of $\sigma $ and $\pi $ virtual photons,
while the last term describes the interaction due to the exchange of virtual
$\sigma $ photons between the atoms situating in different nodes. Again by
assuming equal number of atoms in the two nodes $N_1 = N_2 = N$, we get the
following matrix representation for the new effective Hamiltonian $\hat
{H}_{eff} $ in the basis of the five states

\begin{equation}
\label{eq14}
\left( {{\begin{array}{*{20}c}
 0 \hfill & 0 \hfill & 0 \hfill & 0 \hfill & 0 \hfill \\
 0 \hfill & {\Omega _s N} \hfill & {\Omega _\sigma N} \hfill & 0 \hfill & 0
\hfill \\
 0 \hfill & {\Omega _\sigma N} \hfill & {\Omega _s N} \hfill & 0 \hfill & 0
\hfill \\
 0 \hfill & 0 \hfill & 0 \hfill & {2\Omega _s N} \hfill & {2\Omega _\sigma
\sqrt {N(N - 1)} } \hfill \\
 0 \hfill & 0 \hfill & 0 \hfill & {2\Omega _\sigma \sqrt {N(N - 1)} } \hfill
& {2\Omega _s (N - 1)} \hfill \\
\end{array} }} \right),
\end{equation}

\noindent
where $\Omega _s = \Omega _\sigma + \Omega _\pi $, $\Omega _\sigma = {\vert
g_\sigma \vert ^2} \mathord{\left/ {\vphantom {{\vert g_\sigma \vert ^2}
\Delta }} \right. \kern-\nulldelimiterspace} \Delta $, $\Omega _\pi = -
{\vert g_\pi \vert ^2} \mathord{\left/ {\vphantom {{\vert g_\pi \vert ^2}
\Delta }} \right. \kern-\nulldelimiterspace} \Delta $.

For the initial state (1.1), the atomic wave function evolves as follows

\begin{align}
\label{eq15a}
 \Psi _2 (t) &= \alpha _1 \alpha _2 \psi _1
\nonumber   \\
& + \exp [ - i\Omega _s
Nt]\{\beta _1 \alpha _2 [\cos (\Omega _\sigma Nt)\psi _2 - i\sin (\Omega
_\sigma Nt)\psi _3 ]
+ \alpha _1 \beta _2 [\cos (\Omega _\sigma Nt)\psi _3 - i\sin
(\Omega _\sigma Nt)\psi _2 ]\}
\nonumber   \\
& + \exp [ - i\Omega _s (2N - 1)t]\beta _1 \beta _2 \{[\cos (St) -
i\frac{\Omega _s }{S}\sin (St)]\psi _4 - i\frac{2\Omega _\sigma \sqrt {N(N -
1)} }{S}\sin (St)\psi _5 \},
\end{align}

\noindent
where $\mbox{S} = \sqrt {4\Omega _\sigma ^2 N(N - 1) + \Omega _s^2 } $.

We choose the following parameters for the evolution of Eq. (\ref{eq15a}) providing
the dynamical elimination of the state $\psi _5 $:

\begin{align}
\label{eq15b}
1)& \mbox{ } \Omega _{\sigma} Nt = \pi  ( \textstyle {1\over4} + \textstyle{1\over2} \mu + n); \mbox{ } \mu = 0,1; \mbox{ } n = 0,1,...,
\nonumber  \\
2)& \mbox{ } S t = \pi k, \mbox{ } k = \mbox{1,2,...,}
\end{align}

\noindent
that leads to the following entangled state of the nodes

\begin{align}
\label{eq16}
\Psi _2 (t) &= \alpha _1 \alpha _2 \psi _1
\nonumber   \\
&+ ( - 1)^n\textstyle{1 \over {\sqrt 2 }}\exp [ - i\Omega _s Nt]\{[( - 1)^\mu \beta _1 \alpha _2 -
i\alpha _1 \beta _2 ]\psi _2  + [( - 1)^\mu \alpha _1 \beta _2 -
i\beta _1 \alpha _2 ]\psi _3 \}
\nonumber  \\
& + ( - 1)^k\exp [ - i\Omega _s (2N - 1)t]\beta _1 \beta _2 \psi _4 ,
\end{align}

\noindent
where $\Omega _s $ is determined by the two conditions (\ref{eq15b}). In
particular we write three sets of parameters for possible realizations of CDE
procedure characterized by weaker coupling of atoms with $\sigma $-mode
(n=0,1; $\mu $=0,1):

1) $n = 0, \mu = 0,k = 1: \Omega _\sigma Nt = \pi
/ 4, St = \pi \rightarrow | \Omega _s | t = \sqrt 3 \pi, \frac{| \Omega _s | }{\Omega _\sigma N} = 4 \sqrt 3 \approx 6.92;$

2) $n = 0, \mu = 1,k = 2: \Omega _\sigma Nt =3\pi / 4,St = 2\pi \rightarrow |\Omega _s| t = \sqrt 7 \pi , \frac{| \Omega _s | }{\Omega _\sigma N} =
\frac{4\sqrt 7 }{3} \approx 5.53;$

3) $n = 1,\mu = 0, k = 3: \Omega _\sigma Nt = 5\pi / 4,St = 3\pi \rightarrow | \Omega _s | t = \sqrt 11 \pi , \frac{| \Omega _s | }{\Omega _\sigma N} =\frac{4\sqrt 11 }{5} \approx 2,65$

\noindent
and so on.

Another interesting case occurs for stronger coupling of the atoms with
local $\pi $-modes of the QED cavities when $\vert \Omega _\pi \vert > >
N\Omega _\sigma $. Here, we get a \textit{collective blockade} of state $\psi _5 $ that provides the
following atomic evolution

\begin{align}
\label{eq17}
 \Psi _2 (t)& = \alpha _2 \alpha _3 \psi _1 \mbox{ } + \exp [ - i\Omega _s
Nt]\{\beta _2 \alpha _3 [\cos (\Omega _\sigma Nt)\psi _2 - i\sin (\Omega
_\sigma Nt)\psi _3 ]
\nonumber  \\
& + \alpha _2 \beta _3 [\cos (\Omega _\sigma Nt)\psi _3 - i\sin
(\Omega _\sigma Nt)\psi _2 ]\}
\nonumber   \\
& + \exp [ - i2\Omega _s Nt]\beta _2 \beta _3 \psi _4,
\end{align}

\noindent
yielding the entangled state of the two nodes if only the condition (\ref{eq15b})
is satisfied. So here, we can vary the coupling constant $\Omega _\sigma $
and interaction time t in some possible intervals providing a realization of
general iSWAP gate with arbitrary tunable angle of rotation $\Omega _\sigma
Nt$. \textit{Collective blockade} needs more quality micro-cavities than \textit{collective dynamical elimination technique} but it is more robust being
operative for all necessary temporal durations.

\section{Controlled swap gates}
\label{C_SWAP}

\begin{figure}
  \includegraphics[width=0.6\textwidth,height=0.3\textwidth]{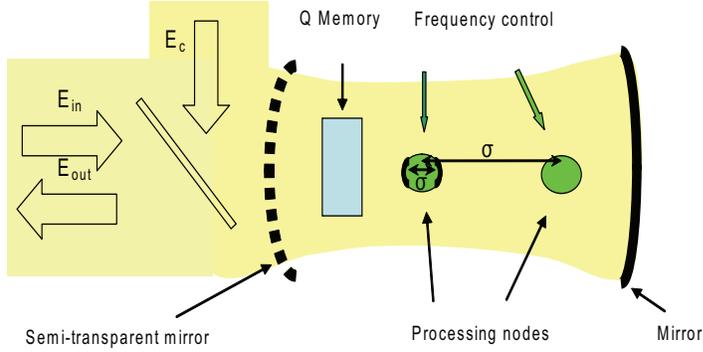}
  \caption{Scheme of control $SWAP$ gate ($CSWAP$) on multi-atomic ensembles in resonator.}
  \label{Figure3}
\end{figure}

Let's consider two atomic ensembles situating in two separate nodes in the
common resonator as shown in Fig.\ref{Figure3}.
With that, one of these nodes has its own
micro-resonator. We can introduce signal and control photons through a beam
splitter into the system. Photons are stored for a time in quantum memory
situating also in common resonator. After absorption of photons by quantum
memory, we raise reflectivity of input-output mirror in order to make
resonator perfect. First, signal photon is transferred from quantum memory
to one of processing nodes and frequency of atomic transitions in processing
nodes is tuned out of resonance with the cavity. Then, we release from
quantum memory the control photon and detune memory from resonance with it.
With that, control photon can not be absorbed by the memory and processing
nodes or released from the cavity at these conditions.

Using Hamiltonian (\ref{eq9}) and states $\psi _1 = \left| 0 \right\rangle _1
\left| 0 \right\rangle _2 $, $\psi _2 = \left| 1 \right\rangle _1 \left| 0
\right\rangle _2 $, $\psi _3 = \left| 0 \right\rangle _1 \left| 1
\right\rangle _2 $, $\psi _4 = \left| 1 \right\rangle _1 \left| 1
\right\rangle _2 $, $\psi _5 = \left| 2 \right\rangle _1 \left| 0
\right\rangle _2 $ and $\psi _6 = \left| 0 \right\rangle _1 \left| 2
\right\rangle _2 $, we get for total wave function

\begin{equation}
\label{eq18}
\psi \left( t \right) = c_1 \left( t \right)\psi _1  + c_2
\left( t \right)\psi _2  + c_3 \left( t \right)\psi _3
+ c_4 \left( t \right)\psi _4  + c_5 \left(
t \right)\psi _5  + c_6 \left( t \right)\psi _6,
\end{equation}

\noindent
the following Schr\"{o}dinger equation

\begin{align}
\label{eq19}
 \frac{d\psi }{dt} & = i\textstyle{N \over {2}} \left\{ \omega_1 + \omega_2 + 2n\left( \Omega _1 + \Omega_2  \right) \right\} c_1 \psi _1
\nonumber  \\
&+ i\left\{ {\left(\textstyle{N \over {2}}-1 \right)\left( {\omega_1 + 2n\Omega_1 } \right) +
\textstyle{N \over {2}}\left( {\omega_2 + 2n\Omega_2 } \right)}
- i N \Omega_{1} \right\}c_2 \psi_2 - iN\Omega_{s} c_{3} \psi _2
 \nonumber   \\
 &- iN \Omega_{s} c_{2} \psi _3 + i\left\{ {\textstyle{N \over {2}}\left( {\omega_1 +
2n\Omega_1 } \right) + \left( {\frac{N}{2} - 1} \right)\left( {\omega_2 +
2n\Omega_2 } \right)} - N\Omega_{2} \right\} c_{3} \psi_3
\nonumber \\
& + i \left\{ \left( {\textstyle{N \over {2}} - 1} \right)\left( {\omega_1 + \omega_2 + 2n\left(
{\Omega_1 + \Omega_2 } \right)} \right) - N\left( {\Omega_1 + \Omega_2 } \right)\right\} c_{4} \psi _4
- i\Omega_s \sqrt {2N\left( {N - 1} \right)}\left( c_{5}+c_6 \right) \psi_4
\nonumber   \\
& - i\Omega_s \sqrt {2N\left( {N - 1} \right)}c_{4} \psi _5
+ i\left( {\textstyle{N \over {2}} - 2} \right)\left( {\omega_1 + 2n\Omega_1 }
\right) c_{5} \psi _5
\nonumber  \\
& - i\Omega_s \sqrt {2N\left( {N - 1} \right)}c_{4} \psi_6
+ i\left( {\textstyle{N \over {2}} - 2} \right)\left( {\omega_2 + 2n\Omega_2 } \right) c_{6} \psi_6,
\end{align}

\noindent
where we have assumed the mode field is in the state with definite number $n$ of photons, $\Omega _1 =\textstyle{\left| {g_{k_0 }^{\left( 1 \right)} } \right|^2 \over {\hbar ^2\Delta}}$, $\Omega _2 =\textstyle{\left| {g_{k_0 }^{\left( 2 \right)} } \right|^2 \over {\hbar ^2\Delta}}$ and
$\Omega _s =\textstyle{ {g_{k_0 }^{\left( 1 \right)} g_{k_0 }^{\left( 2 \right)} }  \over {\hbar ^2\Delta}}$
If $g_{k_0 }^{\left( 2 \right)} \ll g_{k_0 }^{\left( 1 \right)} $. Below we are interested in the case when $\Omega _2 \cong 0$ (second node is characterized by lower quality factor in comparison with the first node factor) and equations for $c_2 $ and $c_3 $ can be
written as

\begin{equation}
\label{eq23}
\frac{dc_2 }{dt} = i\left\{ {\left( {\frac{N}{2} - 1} \right)\left( {\omega
_1 + 2n\Omega _1 } \right) + \frac{N}{2}\omega _2 - N\Omega _1 } \right\}c_2
- iN\Omega _s c_3 ,
\end{equation}

\begin{equation}
\label{eq24}
\frac{dc_3 }{dt} = i\left\{ {\frac{N}{2}\left( {\omega _1 + 2n\Omega _1 }
\right) + \left( {\frac{N}{2} - 1} \right)\omega _2 } \right\}c_3 - iN\Omega
_s c_2,
\end{equation}

\noindent
or

\begin{equation}
\label{eq25}
\frac{dc_2 }{dt} = \frac{i}{\hbar }E_2 c_2 - iN\Omega _s c_3 ,
\end{equation}

\begin{equation}
\label{eq26}
\frac{dc_3 }{dt} = \frac{i}{\hbar }E_3 c_3 - iN\Omega _s c_2,
\end{equation}

\noindent
where $E_2 = \left( {\textstyle{N \over {2}}- 1} \right)\left( {\omega _1 + 2n\Omega _1 }
\right) + \textstyle{N \over {2}}\omega _2 - N\Omega _1 $,
and $ E_3 = \textstyle{N \over {2}}\left( {\omega _1 + 2n\Omega _1 } \right) + \left(
{\textstyle{N \over {2}} - 1} \right)\omega _2$ . The
Equations (\ref{eq25}), (\ref{eq26}) give the following second order equation

\begin{equation}
\label{eq29}
\frac{d^2c_3 }{dt^2} - \frac{i}{\hbar }\left( {E_2 + E_3 } \right)\frac{dc_3
}{dt} - \left( {\frac{E_2 E_3 }{\hbar ^2} - N^2\Omega _s^2 } \right)c_3 = 0,
\end{equation}

\noindent
with a solution

\begin{equation}
\label{eq30}
c_3 = C_1 e^{i r_1 t} + C_2 e^{i r_2 t},
\end{equation}

\noindent
where

\begin{align}
\label{eq31}
 r_{1,2}& = \frac{1}{2\hbar }\left\{ {\left[ {\left( {\textstyle{N \over {2}} - 1}
\right)\left( {\omega _1 + 2n\Omega _1 } \right) + \textstyle{N \over {2}}\omega _2 -
N\Omega _1 } \right] + \left[ {\textstyle{N \over {2}}\left( {\omega _1 + 2n\Omega _1
} \right) + \left( {\textstyle{N \over {2}}- 1} \right)\omega _2 } \right]} \right\}
\nonumber \\
 & \pm \sqrt {\textstyle{1 \over {4}}\left( {\omega _1 - \omega _2 + N\Omega _1 +
2n\Omega _1 } \right)^2 + N^2\Omega _s^2 }.
\end{align}

With the initial conditions $c_2 = 1$ and $c_3 = 0$, we have

\begin{equation}
\label{eq32}
C_1 = - C_2 = - \frac{N\Omega _s }{\sqrt {\left( {\omega _1 - \omega _2 +
N\Omega _1 + 2n\Omega _1 } \right)^2 + N^2\Omega _s^2 } },
\end{equation}

\noindent
that simplifies at $\omega _1 - \omega _2 + N\Omega _1 = 0$ to

\begin{equation}
\label{eq34}
C_1 = - C_2 = - \frac{N\Omega _s }{\sqrt {N^2\Omega _s^2 + 4n^2\Omega _{_1}^2 } }.
\end{equation}

We see that if $n = 1$ at $2\Omega _{1} \gg N\Omega_s $ we have $C_1 =
C_2 = c_3 \cong 0$ and if $n = 0$ we have $C_1 = - 1$,  $C_2 = 1$ and in the
first case no swap occurs and in the second case we have swapping solution

\begin{equation}
\label{eq35}
c_2 = e^{\frac{i}{2\hbar }\left\{ {\left( {\frac{N}{2} - 1} \right)\omega _1
+ \frac{N}{2}\omega _2 - N\Omega _1 } \right\}t}\cos \left( {N\Omega _s t}
\right),
\end{equation}

\noindent
and

\begin{equation}
\label{eq36}
c_3 = - ie^{\frac{i}{2\hbar }\left\{ {\left( {\frac{N}{2} - 1} \right)\omega
_1 + \frac{N}{2}\omega _2 - N\Omega _1 } \right\}t}\sin \left( {N\Omega _s t} \right).
\end{equation}

\noindent
In the first case no swap occurs and in the second case we have swapping solution.

\section{Summary}

So, we have considered $iSWAP$, $\sqrt {iSWAP} $ and  $CSWAP$ gates. $iSWAP$ gate
can be used for efficient transfer of qubit between various nodes of quantum
computer. $\sqrt {iSWAP} $ gate which entangles the two qubits provides a
complete set of universal quantum gates together with single qubit
operations. Here, we note that the single qubit gates can be performed by
transfer the atomic qubit to photonic qubit in waveguide where it can be
rotated on arbitrary angle by usual optical tools \cite {Kok}. We can also return
the qubit back to QM node on demand as it has been shown above. Another
possibility to implement the single qubit gates is to transfer it to the
node with single resonant atom which state can be controlled by external
classical field \cite{Nielsen2000}. Also we can mark the principle possibility to exploit
the collective blockade mechanism for realization of the single qubit gate
similar to approach developed for usual blockade mechanism \cite {Saffman2010} and exploitation of Raman transition between the collective atomic states \cite{Shahriar2007}. Fast $CSWAP$ gate can be used for efficient realization of promising quantum algorithm of
fingerprinting \cite{Buhrman}.
The proposed protocols of two and three-qubit gates also make a creation of large scale universal quantum computer more feasible with the multi-atomic encoding of the single qubit states.

\bibliographystyle{eptcs}

\end{document}